\documentclass[english,aps,superscriptaddress,prl,twocolumn]{revtex4-2}
\usepackage{amsmath}
\usepackage[encapsulated]{CJK}

\usepackage{newtxmath}
\usepackage[T1]{fontenc}
\usepackage{textcomp}
\usepackage[utf8]{inputenc}
\usepackage{color}
\usepackage{pifont}
\usepackage{graphicx}
\makeatletter

\usepackage[colorlinks=true,linkcolor=blue,citecolor=blue,urlcolor=blue]{hyperref}
\AtBeginDocument{%
  \addto\captionsenglish{%
    \renewcommand{\figurename}{FIG.}%
  }%
}

\makeatother

\usepackage{babel}
\begin{document}
\renewcommand{\figurename}{FIG.}
\title{First-Order Topological FFLO Transition and Superconducting Diode
Sign Reversal in Altermagnetic Nanowires}
\author{Bo Fu}
\email{fubo@gbu.edu.cn}

\affiliation{School of Sciences, Great Bay University, Dongguan, China}
\author{Kaizhi Bai}
\affiliation{Department of Physics, The University of Hong Kong, Pokfulam Road,
Hong Kong, China}
\author{Chang-An Li}
\email{changanli@ustc.edu.cn}

\affiliation{Hefei National Laboratory, Hefei, Anhui 230088, China}
\affiliation{School of Emerging Technology, University of Science and Technology
of China, Hefei, Anhui 230026, China}
\author{Shun-Qing Shen}
\affiliation{Department of Physics, The University of Hong Kong, Pokfulam Road,
Hong Kong, China}
\date{\today}
\begin{abstract}
Fulde-Ferrell-Larkin-Ovchinnikov (FFLO) state conventionally emerges
via a second-order phase transition driven by finite magnetization.
Here we show that a spin-orbit-coupled nanowire proximitized to $d$-wave
altermagnets---with zero net magnetization---can realize topological
FFLO states through a first-order transition, marked by a sharp sign-reversing
superconducting diode effect. The altermagnetic field generates band-resolved
competing pairing channels, giving rise to a double-valley free energy
landscape whose global minimum switches discontinuously. It consequently
leads to a first-order topological FFLO transition with simultaneous
jumps in the Cooper pairing amplitude and finite center-of-mass momentum.
Remarkably, this discontinuous topological reconfiguration substantially
enhances the diode efficiency and drives a characteristic sharp sign
reversal across the transition. The mechanism of such exotic phenomena
is captured by Ginzburg--Landau theory. Our results provide a field-free
altermagnetic route to topological FFLO states and identify their
direct transport fingerprint.
\end{abstract}
\maketitle

\paragraph{\textcolor{blue}{Introduction.--}}

The interplay between superconductivity and magnetism continues to
be a fertile ground for discovering new quantum phases \citep{ginzburg1957ferromagnetic,chubukov2012pairing,gor1964ferromagnetism,mathur1998magnetically}
and exotic transport phenomena \citep{buzdin2005proximity,linder2015superconducting,bergeret2005odd}.
A paradigmatic example is the Fulde-Ferrell-Larkin-Ovchinnikov (FFLO)
state, in which the spin imbalance from finite magnetization forces
spin-singlet Cooper pairs to condense at finite center-of-mass momentum,
producing a spatially modulated order parameter \begin{CJK}{GB}{}\citep{fulde1964superconductivity,larkin1965nonuniform,barzykin2002inhomogeneous,agterberg2020physics,Kinjo22Science}\end{CJK}.
Recently, remarkable progress in two closely related fields open new
opportunities to advance FFLO physics. The first is the emergence
of altermagnets, a distinct class of collinear compensated magnetic
order characterized by highly anisotropic, momentum-dependent spin-split
bands \citep{Naka19NC,Smejkal20SACrystal,ShaoDF21NC,Smejkal22prxc,Libor22prx2,BaiL24AFM,QianZ25prb,GuMQ25PRL,Song25NRM,Hayami19JPSJ}.
Unlike conventional ferromagnets, altermagnets exhibit vanishing net
magnetization \citep{LiuZY24prl,Brekke23prb,ZHouXD24PRL,Leeb24prl,LeeS24prl,LiuYC24PRL,Krempasky24Nature,MaH21NC,ezawa2024topological,Antoneko25prl,chen2025probing,li2025altermagnetism,Durrnagel25prl,fukaya2025superconducting,CaoJY25prl,zhu2025altermagnetic,LiuLS26prl,Monkman26prx,LiCA26prl},
thus they can influence Cooper pairing without suppressing superconductivity
\citep{Ouassou23PRL,Beenakker23prb,ZhangSB24NC,DiZhu23PRB,Papaj23PRB,SunC23prb,giil2024superconductor,Ghorashi24PRL,lu2025engineering,SunHP25PRB,alam2025proximity,heinsdorf2025proximitizing}.
The second is the prediction of topological FFLO superconductivity,
where finite-momentum Cooper pairing coexists with nontrivial topology
\begin{CJK}{GB}{}\citep{zhang2013topological,cao2014gapless,chen2013inhomogeneous,liu2013topological,qu2013topological}\end{CJK}.
Such phases are predicted to host Majorana bound states and exhibit
nonreciprocal superconducting transport signatures. Yet existing proposals
for topological FFLO states typically rely on finite magnetization
or external Zeeman fields, the very ingredients that altermagnetism
could render unnecessary. To date, however, whether altermagnetism---with
zero net magnetization and momentum-dependent exchange field---can
stabilize and control topological FFLO states remains largely unexplored,
despite some related efforts \citep{ZhangSB24NC,SimB25prb,samanta2025field,liu2025altermagnetism,LiuZ26prb,LiuXJ26arxiv}.
This question is particularly compelling since the altermagnetic exchange
field could qualitatively reshape the pairing landscape and change
the band topology.

\begin{figure}
\includegraphics[width=8.8cm]{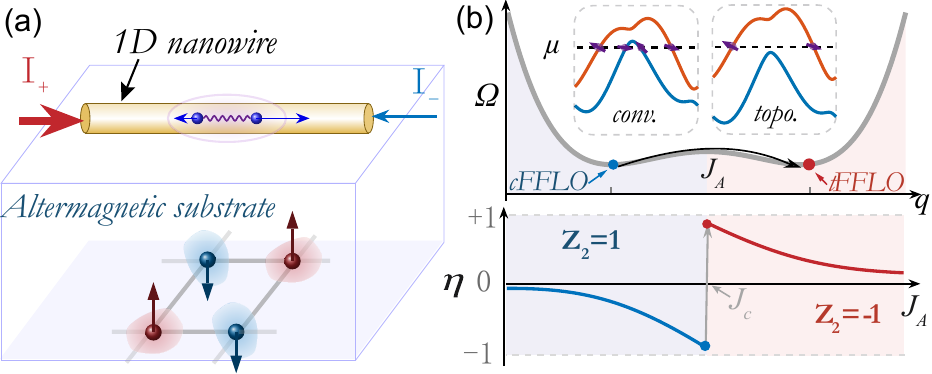}\caption{(a) Schematic of a 1D spin-orbit-coupled nanowire proximitized
to a $d$-wave altermagnetic substrate. The emergence of FFLO states
stabilizes finite-momentum pairing and an intrinsically nonreciprocal
superconducting state with unequal critical currents $I_{+}\protect\neq I_{-}$.
(b) Illustration of the first-order topological FFLO transition driven
by altermagnetism. The proximity-induced altermagnetic filed tilts
energy bands of the spin-orbit-coupled nanowire and reshape their
spin texture, producing band-resolved competing pairing channels (two
insets). Upon tuning $J_{A}$, the system undergoes a first-order
topological phase transition between conventional ($\mathbb{Z}_{2}=+1$)
and topological ($\mathbb{Z}_{2}=-1$) FFLO states. The transition
is marked by a discontinuous jump in the Cooper-pair momentum $q$
and pairing amplitude $\Delta$ in the two-valley free-energy $\Omega$
landscape (upper panel). It is further directly probed by the superconducting
diode coefficient $\eta$ that is strongly enhanced and reverses sign
across the transition (lower panel). \protect\label{fig:=000020Schematic_diagram}}
\end{figure}

In this work, we investigate the altermagnetism-driven topological
FFLO transition and its consequent superconducting transport signatures.
We consider a spin-orbit-coupled nanowire with Hubbard interactions
proximitized to a $d$-wave altermagnet substrate, as illustrated
in Fig. \ref{fig:=000020Schematic_diagram}(a). The proximity-induced
altermagnetic exchange field in the nanowire breaks mirror/inversion
and time-reversal symmetries while preserving zero net magnetization.
By tilting the spin-split bands and reshaping their spin textures,
this momentum-dependent field generates competing band-resolved pairing
channels. Using self-consistent Bogoliubov-de Gennes theory, we establish
the existence of topological FFLO states that carry both finite center-of-mass
momentum and a nontrivial $\mathbb{Z}_{2}$ invariant. Strikingly,
the transition from conventional FFLO ($\mathbb{Z}_{2}=+1$) to topological
FFLO ($\mathbb{Z}_{2}=-1$) is first order, with simultaneous discontinuous
jumps in the pairing amplitude and finite Cooper-pair momentum. This
first-order topological reconfiguration directly imprints on the emergent
superconducting diode effect: the diode efficiency $\eta$ is strongly
enhanced and reverses sign sharply across the phase boundary {[}see
Fig. \ref{fig:=000020Schematic_diagram}(b){]}. We elucidate the mechanism
through a Ginzburg-Landau analysis, which reveals that band-resolved
pairing channels reshape the free-energy landscape into a double-valley
structure, with two local minima corresponding to the conventional
and topological FFLO states {[}see two insets in Fig. \ref{fig:=000020Schematic_diagram}(b){]}.
Tuning the altermagnetic exchange switches the global minimum discontinuously
between these valleys, thereby underpinning both the first-order topological
phase transition and the highly nonreciprocal supercurrent. Finally,
we demonstrate the robustness of these phenomena at finite temperature
and discuss their experimental feasibility.

\paragraph{\textcolor{blue}{Model and method.--}}

\begin{figure}
\includegraphics[width=8.8cm]{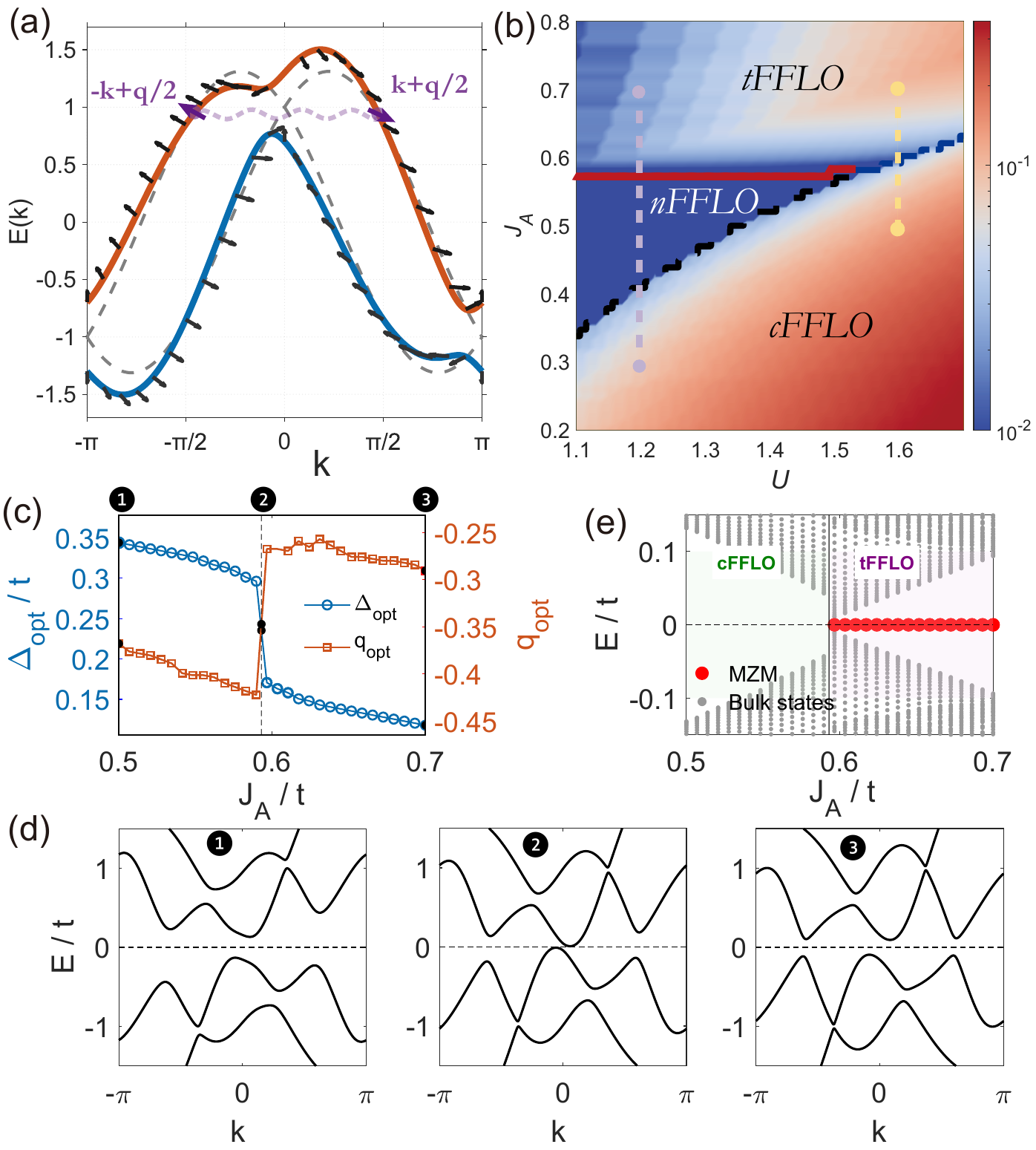}

\caption{(a) Normal-state band structure $E(k)$: the dashed curve, mirror-symmetric
spectrum at $J_{A}=0$ (SOC only); solid curves, tilted, asymmetric
bands at finite $J_{A}$. Inset: schematic finite-$q$ pairing between
shifted Fermi points leading to FFLO order. (b) Phase diagram in the
($U,J_{A}$) plane at $\mu=0.55t$. Colors indicate the superconducting
mini-gap (blue: nodal FFLO; red: gapped FFLO). Black and blue dashed
lines mark first-order boundaries between cFFLO--nFFLO and cFFLO--tFFLO,
respectively; the solid red line marks the (continuous) transition
between nFFLO and tFFLO. (c) Self-consistent optimal pairing amplitude
$\Delta$ (blue circles, left axis) and Cooper-pair momentum $q$
(red squares, right axis) as functions of $J_{A}$ along the cut indicated
by the green line in panel (b), showing simultaneous jumps at the
first-order transition (vertical dashed line). (d) Bulk BdG band structures
for three representative $J_{A}$ values before (\ding{182}), at (\ding{183}),
and after (\ding{184}) the transition. (e) Corresponding open-boundary
BdG spectrum versus $J_{A}$ along the same cut. Filled red symbols
indicate in-gap states (Majorana zero modes) that appear or disappear
when the ground state switches between cFFLO and tFFLO. All energies
are in units of $t=1$ with with $\alpha_{y}=\alpha_{z}=0.6t$, and
$U=1.5t$ unless otherwise specified.\protect\label{fig:1st_order_TPT}}
\end{figure}

We consider a one-dimensional (1D) spin-orbit-coupled nanowire with
attractive Hubbard interactions proximitized to $d$-wave altermagnets
substrate {[}see Fig. \ref{fig:=000020Schematic_diagram}(a){]}. The
normal-state Bloch Hamiltonian reads 
\begin{equation}
H(k)=\xi(k)+\alpha_{y}\sin k\sigma_{y}+(\alpha_{z}\sin k+J_{A}\cos k)\sigma_{z},
\end{equation}
where $\xi(k)=t\cos k$ with $k$ the wave vector and $t$ the hopping
amplitude. $\sigma_{x,y,z}$ are Pauli matrices in spin space. The
parameters $\alpha_{y}$ and $\alpha_{z}$ denote spin-flip Rashba
and spin-conserving Dresslhaus spin-orbit coupling (SOC) strength
\citep{bychkov1984oscillatory,dresselhaus1955spin}, respectively,
while $J_{A}$ is the altermagnetic exchange strength induced by proximity
to the substrate \citep{Ghorashi24PRL,zhu2025altermagnetic}. The
altermagnetic term $J_{A}\cos k\sigma_{z}$ generates momentum-dependent
spin splitting that simultaneously breaks time-reversal symmetry and
mirror symmetry $M_{x}$: $k\rightarrow-k,\boldsymbol{\sigma}\to(\sigma_{x},-\sigma_{y},-\sigma_{z})$.
Consequently, as shown in Fig. \ref{fig:1st_order_TPT}(a), the mirror-symmetric
Rashba-only bands presented at $J_{A}=0$ (gray dashed curve) acquire
an asymmetric, momentum-tilted and spin-split dispersion at finite
$J_{A}$. The asymmetric \textquotedblleft tilted\textquotedblright{}
band structure (red and blue lines), with energy gaps shifted away
from high-symmetry points, changes the Fermi momenta and reshape the
spin textures at the Fermi points. As a result, the favored pairing
phase space is modified, biasing distinct center-of-mass pairing momenta
$q$ for different bands upon superconductivity is introduced.

To stabilize finite-momentum singlet paring, we include an on-site
attractive Hubbard interaction \citep{chakraborty2025perfect,sim2025pair}
\begin{equation}
H_{I}=-\frac{U}{N_{c}}\sum_{k,k^{\prime},q}c_{k+q/2,\uparrow}^{\dagger}c_{-k+q/2,\downarrow}^{\dagger}c_{-k^{\prime}+q/2,\downarrow}c_{k^{\prime}+q/2,\uparrow},
\end{equation}
where $c_{k\sigma}^{\dagger}$ ($c_{k\sigma}$) creates (annihilates)
an electron with momentum $k$ and spin $\sigma$, $U>0$ is the interaction
strength, and $N_{c}$ is the number of lattice sites. Performing
a mean-field decoupling in the FFLO channel with a uniform center-of-mass
momentum \citep{Fu2026SM}, the order parameter is $\hat{\Delta}(q)=\frac{U}{N_{c}}\sum_{k}c_{-k+\frac{q}{2}\downarrow}c_{k+\frac{q}{2}\uparrow}.$
The resulting Bogoliubov-de Gennes (BdG) Hamiltonian in the Nambu
basis $\Psi_{k}=\{c_{k+q/2,\uparrow},c_{k+q/2,\downarrow},c_{-k+q/2,\uparrow}^{\dagger},c_{-k+q/2,\downarrow}^{\dagger}\}$
takes the form \begin{CJK}{GB}{}
\begin{equation}
\mathcal{H}(k)=\left(\begin{array}{cc}
H(k+q/2)-\mu & -i\sigma_{y}\Delta(q)\\
i\sigma_{y}\Delta^{*}(q) & -H^{*}(-k+q/2)+\mu
\end{array}\right),
\end{equation}
\end{CJK}where $\mu$ is the chemical potential.
Diagonalization yields quasiparticle eigenvalues $E_{kq}^{\zeta\lambda}$,
where $\zeta$ and $\lambda$ label band and particle-hole indices,
respectively. At zero temperature, the free-energy density per site
is\begin{CJK}{GB}{}
\begin{align}
F(\Delta,q) & =\frac{1}{2N_{c}}\sum_{k,\zeta,\lambda}E_{kq}^{\zeta\lambda}\Theta(-E_{kq}^{\zeta\lambda})+\frac{\Delta^{2}}{U}-\mu,
\end{align}
\end{CJK}where $\Theta$ is the Heaviside step
function. For each parameter set $(\mu,J_{A},U)$, the equilibrium
pairing state is determined by self-consistently minimizing the condensation
energy $\Omega(\Delta,q)=F(\Delta,q)-F(0,q)$ over the space $(\Delta,q)$
to locate the global and local minima \citep{santra2026superconducting,zhang2023kramers}.

The 1D BdG Hamiltonian admits an antiunitary particle-hole symmetry
$\mathcal{P}\mathcal{H}^{*}(k)\mathcal{P}^{-1}=-\mathcal{H}(-k)$
with $\mathcal{P}=\tau_{x}$ acting in the Nambu space, placing the
system in symmetry class D of the Altland--Zirnbauer tenfold classification
\citep{ChiuCK16rmp}. When the quasiparticle spectrum is fully gapped,
the topology is characterized by a $\mathbb{Z}_{2}$ invariant, constructed
from the Pfaffian of the Hamiltonian in the Majorana basis evaluated
at the momenta $k=0,\pi$ \citep{Fu2026SM}. For the Majorana-basis
Hamiltonian $\mathcal{H}^{\prime}(k)$, the topological invariant
is given by the Pfaffian criterion
\begin{alignat}{1}
\mathrm{Pf}[\mathcal{H}^{\prime}(0/\pi)] & =[\mu\pm t\cos(q/2)]^{2}+\Delta^{2}-\alpha_{y}^{2}\sin^{2}(q/2)\nonumber \\
 & -[J_{A}\cos(q/2)+\alpha_{z}\sin(q/2)]^{2}.\label{eq:Pfaffian}
\end{alignat}

\paragraph*{\textcolor{blue}{First-order topological FFLO transition.--}}

We first map out the phase diagram of different FFLO states and then
focus on the exotic first-oder topological FFLO transition. Minimizing
$\Omega(\Delta,q)$ over the parameter space $(U,J_{A})$ yields the
phase diagram as shown in Fig. \ref{fig:1st_order_TPT}(b). Three
distinct FFLO phases emerge: (i) a gapped conventional FFLO (cFFLO)
with trivial invariant $\mathbb{Z}_{2}=+1$; (ii) a nodal FFLO (nFFLO)
with Bogoliubov Fermi surfaces (blue region) \citep{brydon2018bogoliubov,fu2025altermagnetism,fukaya2025crossed,pal2024identifying,wu2025intra,setty2020topological,timm2021symmetry,oh2021using,yuan2022supercurrent,mo2025coexistence};
and (iii) a gapped topological FFLO (tFFLO) with $\mathbb{Z}_{2}=-1$.
Notably, the cFFLO--nFFLO and cFFLO--tFFLO transitions are first-order
(dashed boundary lines), while the nFFLO--tFFLO transition is a continuous
second-order transition (red solid boundary line).

To characterize the nature of the cFFLO--tFFLO boundary, we follow
the yellow line cut in Fig. \ref{fig:1st_order_TPT}(b) at fixed $U=1.6t$
while varying $J_{A}$. As shown in Fig. \ref{fig:1st_order_TPT}(c),
both the optimal order parameter $\Delta$ and pairing finite momentum
$q$ jump discontinuously at a critical value $J_{c}\approx0.596t$.
It demonstrates the first-order phase boundaries separating the nFFLO--tFFLO
states \citep{beyer2012angle,barbarino2019first,jurivcic2017first,agarwal2023first,yonezawa2013first}.
Correspondingly, the bulk quasiparticle spectrum of the periodic BdG
Hamiltonian still experiences a gap closing and reopening process,
yet in a manner qualitatively different from conventional topological
transitions {[}Fig. \ref{fig:1st_order_TPT}(d){]}. We find that the
bulk gap collapses without a stable band-touching point in the low-energy
spectrum during such a first-order transition {[}see point \ding{183}
in Fig. \ref{fig:1st_order_TPT}(d){]}. This discontinuous closing
allows the system to change its topological invariant via a level
crossing between competing gapped minima at a fixed momentum. This
mechanism permits hysteretic behavior and sharp transport signatures
near the transition. The topological character of the switch is confirmed
by open-boundary spectra across the transition {[}Fig. \ref{fig:1st_order_TPT}(e){]}.
It demonstrates the appearance of Majorana zero modes (MZMs) upon
entering the tFFLO phase, in full agreement with the Pfaffian criterion
of Eq.\ (\ref{eq:Pfaffian}).

\begin{figure}
\includegraphics[width=8.8cm]{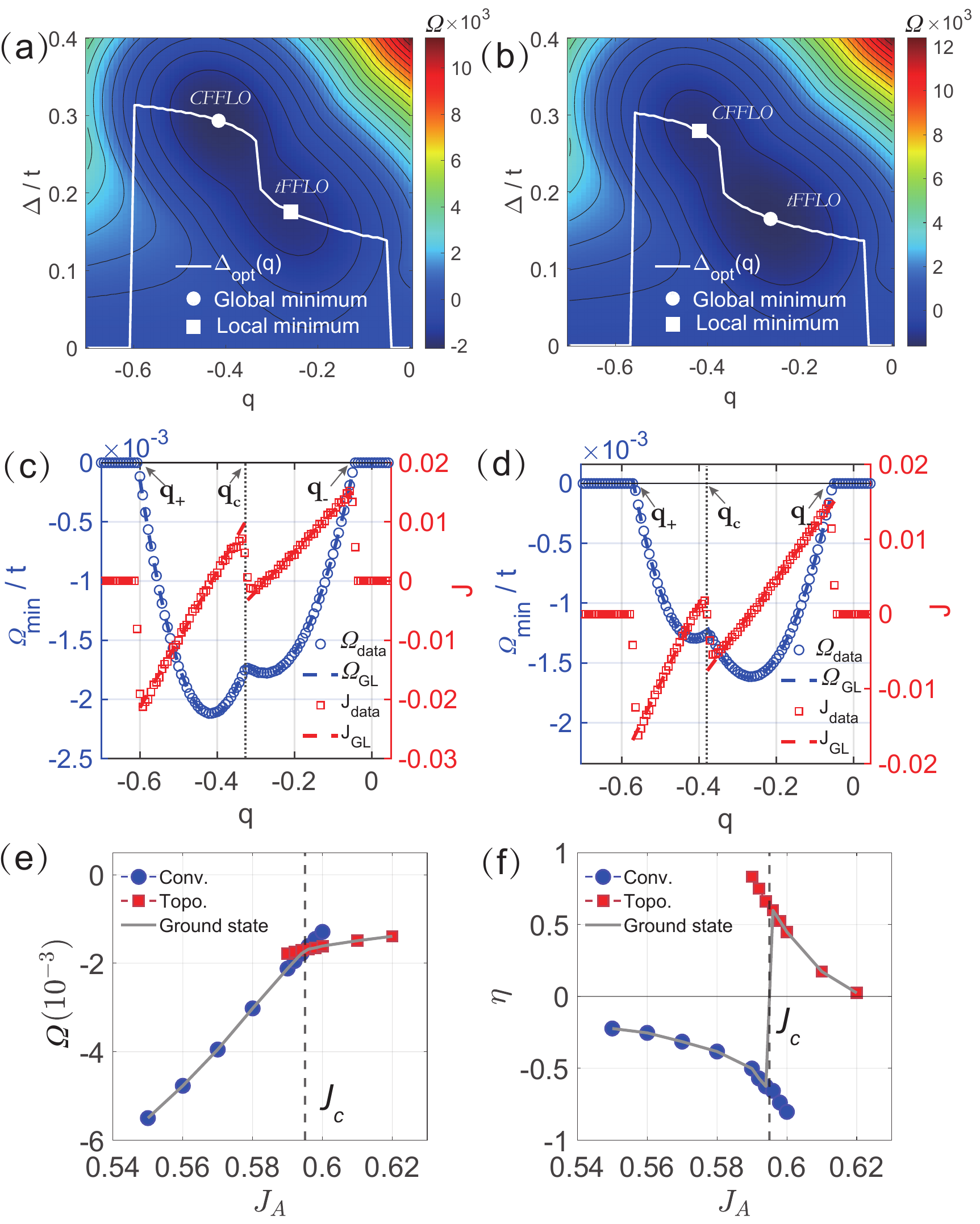}

\caption{(a,b) Representative condensation-energy density $\Omega(\Delta,q)$
at $U=1.5t$ for two altermagnetic strengths $J_{A}=0.59t$ and $J_{A}=0.60t$,
just below and above the critical value $J_{c}\approx0.595t$. Both
panels display two minima --- cFFLO ($\mathbb{Z}_{2}=+1$) and tFFLO
($\mathbb{Z}_{2}=-1$) --- separated by a saddle point. White markers
denote global and local extrema; the white curve traces the optimal
$\Delta(q)$ that minimizes $\Omega$ at each $q$. (c,d) Corresponding
cuts showing $\Omega(q)$ (blue) and the supercurrent density $J(q)$
(red) along the optimal $\Delta(q)$ line. The shift of the global
minimum between (c) and (d) signals the first-order transition. Open
symbols represent the numerical results, while dashed lines denote
the corresponding Ginzburg-Landau theory. (e) Condensation energies
of the cFFLO (blue circles) and tFFLO (red squares) minima versus
$J_{A}$. The clear level crossing marks the discontinuous ground-state
switch. (f) Corresponding superconducting-diode efficiency $\eta$
versus $J_{A}$, exhibiting a sharp sign reversal concurrent with
the first-order transition.\protect\label{fig:free_energy_diode_effect}}
\end{figure}

The free-energy landscapes in Figs. \ref{fig:free_energy_diode_effect}(a)
and \ref{fig:free_energy_diode_effect}(b) directly reveal the energetic
mechanism underlying the observed behavior. On either side of $J_{c}$,
$\Omega(\Delta,q)$ exhibits two distinct minima, a deep global minimum
and a shallower local one, separated by a saddle point. The transition
occurs when their relative depths invert: the global minimum switches
discontinuously from one to the other. For $J_{A}<J_{c}$, the global
minimum (white circle) corresponds to the cFFLO. For $J_{A}>J_{c}$,
the relative depths of minima invert and the ground state shifts to
the tFFLO pairing phase. Since the two separated minima are not adiabatically
connected, this inversion produces the abrupt jumps in order parameter
$\Delta$ and finite momentum $q$.

Explicitly, the blue circle lines in Figs. \ref{fig:free_energy_diode_effect}(c)
and \ref{fig:free_energy_diode_effect}(d) show cuts of $\Omega(q)$
along the optimal trajectory $\Delta_{\text{opt}}(q)$ corresponding
to the white lines in Figs. \ref{fig:free_energy_diode_effect}(a)
and \ref{fig:free_energy_diode_effect}(b), which transparently capture
level crossing as $J_{A}$ passes through $J_{c}$. This level crossing
has a direct band-structure origin. At small $J_{A}$, the chemical
potential $\mu$ intersects two spin-split bands as shown in Fig.\,\ref{fig:1st_order_TPT}(a).
Each band imposes its own preferred $(\Delta,q)$ through its Fermi
momentum and spin texture, generating competing pairing channels that
stabilize the cFFLO. At large $J_{A}$, in contrast, the chemical
potential can intersect effectively only one single helical band within
a gap window. As a result, the unique optimally $(\Delta,q)$ fully
gaps the Fermi surface and stabilizes the tFFLO states. Continuously
tuning $J_{A}$ therefore shifts the energetic balance between these
two condensates, driving a first-order transition accompanied by abrupt
changes in $\Delta$, $q$, and the Pfaffian sign. Figure \ref{fig:free_energy_diode_effect}(e)
tracks the corresponding ground-state energies. It further makes clear
that the transition is driven by a level crossing between two competing
condensates. Notably, the chemical potential --- which can be easily
tuned by a gate voltage --- similarly serves as a control parameter
for the phase transition, offering an complementary knob to the altermagnetism
mechanisms discussed above \citep{Fu2026SM}.

\paragraph*{\textcolor{blue}{Sign-reversing superconducting diode effect.--}}

The FFLO state under inversion symmetry breaking gives rise to a superconducting
diode effect characterized by nonreciprocal supercurrent. This nonreciprocity
is encoded in the current--phase relation $J(q)=2\frac{\partial\Omega}{\partial q}$,
whose asymmetry between positive and negative critical currents defines
the superconducting diode effect \citep{daido2022intrinsic,ilic2022theory,he2022phenomenological,yuan2022supercurrent}.
We quantify this asymmetry through the diode efficiency
\begin{alignat}{1}
\eta & =(|I_{c}^{+}|-|I_{c}^{-}|)/(|I_{c}^{+}|+|I_{c}^{-}|),
\end{alignat}
where $I_{c}^{\pm}$ denote the critical currents in opposite directions.
Evaluating $J(q)$ along the optimal trajectory $\Delta_{\text{opt}}(q)$
{[}red lines in Figs. \ref{fig:free_energy_diode_effect}(c) and \ref{fig:free_energy_diode_effect}(d){]},
we find that the supercurrent profile undergoes an abrupt reversal
coinciding with the shift of the global minimum in the free-energy
landscape $\Omega(\Delta,q)$. Consequently, the diode efficiency
$\eta$ is highly sensitive to the ground-state center-of-mass momentum
and and to the detailed shape of the current--phase relation $J(q)$,
which are both controlled by the underlying free-energy landscape
in the FFLO state. Near the first-order phase boundary, the two nearly
degenerate free-energy minima are associated with distinct values
of finite momentum $q$ and qualitatively different current distributions.
In this regime, small variations in $J_{A}$ can flip which minimum
constitutes the ground state, rendering the critical current strongly
asymmetric, as shown in Fig. \ref{fig:free_energy_diode_effect}(f).
Microscopically, the sensitivity of $\eta$ thus originates from this
double\nobreakdash-valley structure at the first-order transition,
which makes the ground state extremely susceptible to small parameter
shifts between competing FFLO configurations. Around the phase boundary,
the near-degenerate two minima further amplifies both effects, producing
a large $|\eta|$ and, as confirmed in our numerics, a clear sign
reversal of $\eta$ across the transition. Henceforth, the abruptly
changed sign and magnitude of $\eta$ thus provide a experimentally
accessible probe of the underlying first-order topological FFLO transition.

\paragraph{\textcolor{blue}{Ginzburg-Landau theory description.--}}

To elucidate the nature of the observed first-order FFLO transition
and accompanying superconducting diode effect, we construct a Ginzburg-Landau
free energy
\begin{alignat}{1}
\Omega(q,\Delta) & =\alpha(q)\Delta^{2}+\beta(q)\Delta^{4}+\gamma(q)\Delta^{6},
\end{alignat}
where $\beta(q)<0$ and $\gamma(q)>0$ ensure thermodynamic stability
and the existence of a barrier between competing superconducting states
\citep{tinkham2004introduction}. Imposing the stationary condition
$\partial\Omega/\partial\Delta=0$ yields two nontrivial branches
for the pairing amplitude, $\Delta_{\pm}^{2}=(-\beta\pm\sqrt{\beta^{2}-3\alpha\gamma})/3\gamma$,
each corresponding to a distinct valley in the free-energy landscape.
Substituting $\Delta_{\pm}$ back into $\Omega(q,\Delta)$ defines
two competing branches $\Omega_{\pm}(q)=\pm(\sqrt{\beta^{2}-3\alpha\gamma}\mp\beta)(6\alpha\gamma\pm\beta[\sqrt{\beta^{2}-3\alpha\gamma}\mp\beta])/27\gamma^{2}$
(SM). These branches are well approximated by simple quadratics $\Omega_{\pm}=f_{\pm0}+f_{\pm1}q+f_{\pm2}q^{2}$,
as demonstrated in Figs. \ref{fig:free_energy_diode_effect}(c) and
\ref{fig:free_energy_diode_effect}(d). As $q$ varies, the system
undergoes a sharp, nonanalytic crossover from the $\Omega_{+}$ valley
to the $\Omega_{-}$ valley at a critical momentum $q_{c}$. This
transition is fundamentally driven by the competition between the
local extrema $\Omega_{\pm}^{\mathrm{ex}}(q)=f_{\pm0}-f_{\pm1}^{2}/4f_{\pm2}$.
When an external control parameter, e.g., the altermagnetic strength
$J_{A}$, is tuned such that the system is close to the degeneracy
point $\Omega_{+}^{\mathrm{ex}}=\Omega_{-}^{\mathrm{ex}}$, the global
ground state switches discontinuously between two FFLO states with
distinct momenta and gap magnitudes. The transition is therefore genuinely
of first-order nature.

To connect this free-energy structure to transport features, we define
$q_{\pm}$ as the extinction momenta at which the superconducting
pairing vanishes, i.e., $\Omega_{\pm}(q_{\pm})=0$. Within each branch,
the critical current $J_{\pm}(q)=2\partial\Omega_{\pm}/\partial q=2f_{\pm1}+4f_{\pm2}q$
evolves linearly, spanning the intervals $[q_{+},q_{c}]$ and $[q_{c},q_{-}]$,
respectively. This piecewise-linear feature naturally produces strong
nonreciprocity and a sign reversal of the diode effect near the transition.
The corresponding diode efficiency associated with each branch is
\begin{alignat}{1}
\eta_{\pm} & =\pm\frac{f_{\pm1}/f_{\pm2}+(q_{c}+q_{\pm})}{q_{c}-q_{\pm}}.
\end{alignat}
Thus, as the system switch between the two free-energy valleys corresponding
to different FFLO states, the diode efficiency $\eta$ changes sign
abruptly. The Ginzburg-Landau description provides a robust, analytic
mechanism for simultaneous discontinuities in the pairing strength
and in the superconducting diode response \citep{Fu2026SM}. As shown
in Figs. \ref{fig:free_energy_diode_effect}(c-f) , it captures the
numerically observed first-order transition quantitatively.

\paragraph{\textcolor{blue}{Robustness at finite temperature.--}}

\begin{figure}
\includegraphics[width=8cm]{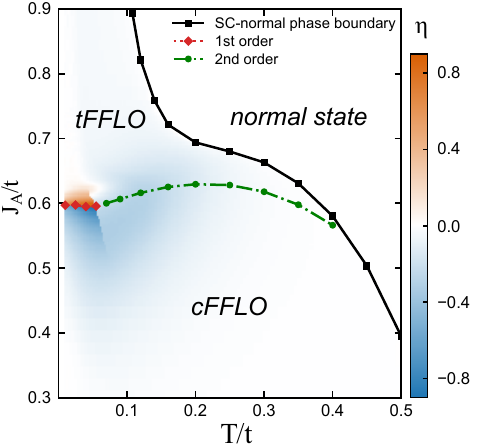}\caption{Finite-temperature phase diagram in the $(J_{A},T)$ plane at $U=1.6t$.
The black dashed curve marks the superconducting--normal boundary;
above it the system is in normal states. Below the boundary (black
solid line), the system resides in either a tFFLO or cFFLO phase.
The color map shows the superconducting diode efficiency $\eta$:
blue (red) indicates negative (positive) $\eta$, white corresponds
to $\eta\approx0$. At low temperatures, the cFFLO--tFFLO boundary
(red dashed line) is first order and $\eta$ changes sign discontinuously;
at higher temperatures this boundary (green dashed line) becomes continuous
and $\eta$ evolves smoothly across it.\protect\label{fig:Temperature}}
\end{figure}

Having established the zero-temperature phase structure, we extend
the analysis to finite temperature. We map out the phase diagram in
the $(J_{A},T)$ plane at fixed $U=1.6t$ as shown in Fig.\,\ref{fig:Temperature}.
Three distinct regions, i.e., the cFFLO at lower part, the tFFLO at
middle part, and the normal states at upper part, are separated by
two phase boundaries (squared and circled lines). Notably, the largest
values of the superconducting diode efficiency $\left|\eta\right|$
concentrate precisely along the cFFLO--tFFLO boundary, consistent
with above analysis. Several features of this phase diagram merit
attention. First, altermagnetism acts to suppress superconductivity:
increasing $J_{A}$ shrinks the superconducting dome and lowers the
critical temperature, producing a narrow, high peak near zero temperature.
The overall phase shape closely resembles the standard phase diagram
of an $s$-wave superconductor in a magnetic field. Second, the character
of the cFFLO--tFFLO boundary evolves with temperature in a manner
that directly reflects the underlying competition between the two
FFLO minima. The color scale encodes the superconducting diode efficiency
$\eta$. At low temperatures, this boundary inherits the first-order
character at $T=0$: $\eta$ exhibits a sharp jump and sign change
when crossing this boundary. As temperature increases, thermal fluctuations
gradually weaken the competition between the two FFLO free-energy
minima, and the boundary evolves into a continuous, second-order line
along which the $\eta$ varies smoothly. Henceforth, the finite-temperature
phase diagram indicate that a sharp sign reversal in $\eta$ signals
the robustness of first-order transition at relatively low temperature
regime, while a smooth variation signals a thermally broadened crossover
at high temperature regime \citep{Fu2026SM}.

\paragraph{\textcolor{blue}{Discussion and conclusions.--}}

In summary, we have demonstrated that a 1D SOC nanowire proximitized
to a $d$-wave altermagnet experiences a first-order topological phase
transition between cFFLO and tFFLO states. The transition is characterized
by simultaneous abrupt jumps in the Cooper pairing amplitude and center-of-mass
momentum. It is accompanied by a strongly enhanced superconducting
diode effect with abrupt sign reversal in $\eta$, providing an accessible
transport probe of the underlying topological transition. The underlying
mechanism, rooted in competing pairing channels and incompatible $q$-preferences
of different bands, suggests feasible routes to realize and control
tFFLO phases via altermagnetism.

A key feature of our proposal is that the topological FFLO state emerges
at zero net magnetization---the altermagnetic exchange splits bands
in momentum space without producing a macroscopic magnetic moment,
thereby circumventing the orbital depairing that limits conventional
Zeeman-driven FFLO. This makes altermagnetic proximity a uniquely
clean route to finite-momentum pairing, decoupled from the flux-induced
complications inherent to applied magnetic field experiments. The
feasible platforms are InAs or InSb nanowires with strong SOC proximitized
to $d$-wave altermagnets such as $\mathrm{Rb_{1-\delta}V_{2}Te_{2}O}$
and $\mathrm{KV_{2}Se_{2}O}$ \citep{JiangB25NP,ZhangF25NP} , where
both $\mu$ and $J_{A}$ are independently tunable via gating and
interface engineering. The predicted hysteresis and sign reversal
of $\eta$ at the cFFLO--tFFLO boundary are unambiguous transport
signatures of the first-order transition, well within reach of existing
diode efficiency measurements \citep{Ando20Nature,Jeon22NatMatt}.
These exotic effects also provide an alternative route to probe Majorana-relevant
physics in topological superconducting nanowires.

\paragraph{\textcolor{blue}{Acknowledgments.--}}

B.F. thanks Zhen Zheng for insightful discussions regarding the FFLO
state. This work is supported by National Natural Science Foundation
of China (Grants No.12504049), Guangdong Province Introduced Innovative
R\&D Team Program (Grant No. 2023QN10X136), Guangdong Basic and Applied
Basic Research Foundation No. 2024A1515010430 and 2023A1515140008),
Research Grants Council, University Grants Committee, Hong Kong under
Grant No. 17301823, and Guangdong Provincial Quantum Science Strategic
initiative (GDZX230005). C.A.L. acknowledges the support from the
Innovation Program for Quantum Science and Technology and the start-up
funding at HFNL (Grant No. QD2022600001).

\bibliographystyle{apsrev4-1-etal-title-url-first}
\bibliography{Refsdata}

\section*{End Matter}

\begin{figure}
\includegraphics[width=7.5cm]{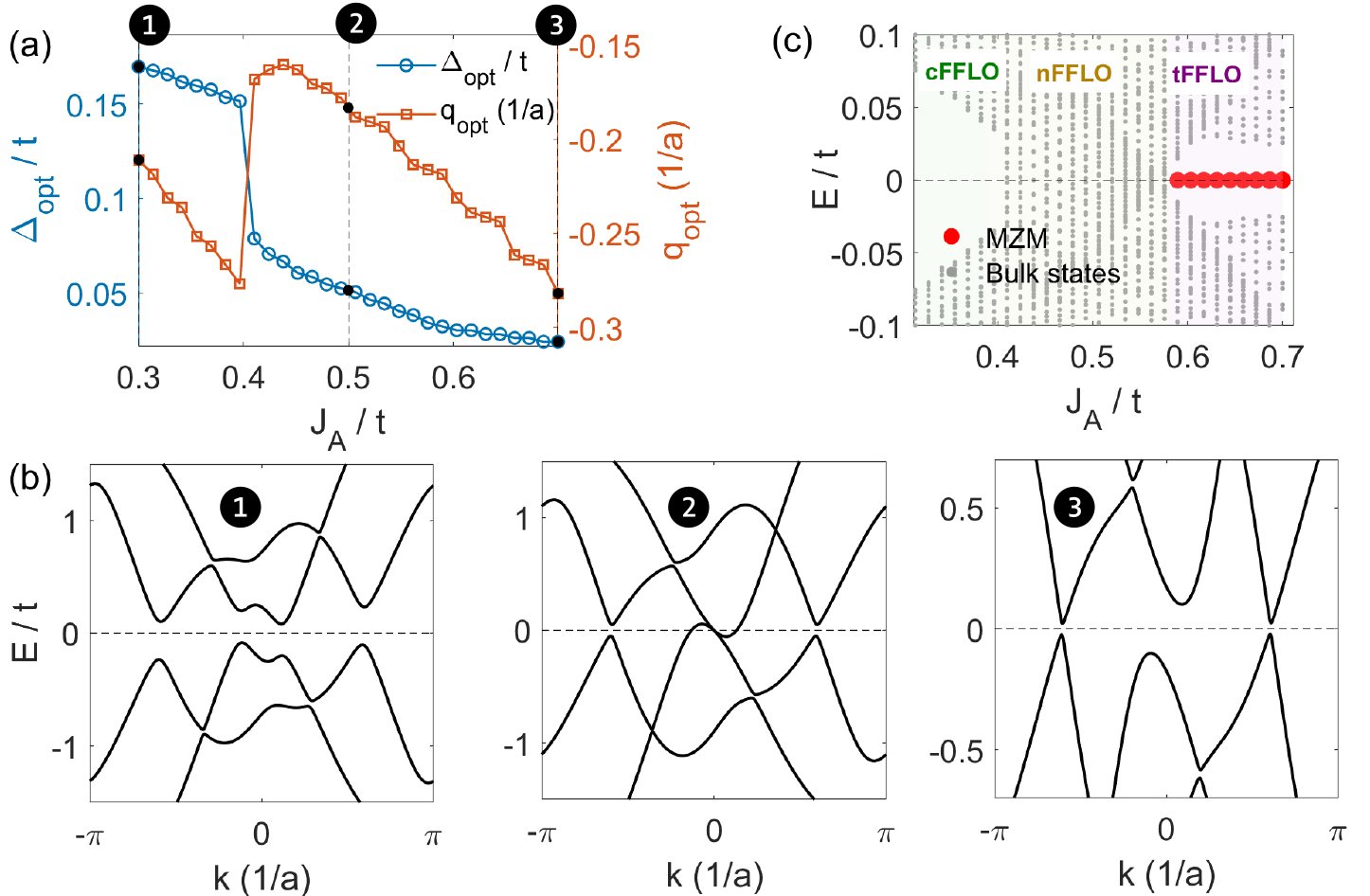}

\caption{Self-consistent evolution across the cFFLO \textrightarrow{} nFFLO
\textrightarrow{} tFFLO sequence {[}gray cut in Fig.\,\,\ref{fig:1st_order_TPT}(b){]}.
(a) Optimal pairing amplitude (blue circles, left axis) and Cooper-pair
momentum (red squares, right axis) versus $J_{A}/t$, showing a first\protect\nobreakdash-order
jump at the cFFLO--nFFLO boundary and a smooth crossover at the nFFLO--tFFLO
boundary. (b) Periodic-boundary BdG spectra at the three $J_{A}$
values indicated by symbols in (a), illustrating the transition from
gapped cFFLO to nodal nFFLO to gapped tFFLO. (c) Corresponding open-boundary
spectra versus $J_{A}/t$, highlighting the appearance of MZMs only
in the tFFLO phase.\protect\label{fig:EndMatt}}
\end{figure}

The main text focuses on the first-order cFFLO--tFFLO transition
and its diode signature. Here we examine the cFFLO \textrightarrow{}
nFFLO \textrightarrow{} tFFLO transition sequence along the gray dashed
cut in Fig.\,\ref{fig:1st_order_TPT}(b). The cFFLO--nFFLO boundary
(black dashed line) is first order: the system jumps from a fully
gapped trivial FFLO state to a nodal FFLO state with Bogoliubov Fermi
surfaces, accompanied by clear discontinuities in the self-consistent
pairing amplitude and Cooper-pair momentum {[}Fig.\,\,\ref{fig:EndMatt}(a){]}.
By contrast, the nFFLO--tFFLO boundary (solid red line) is continuous:
the bulk gap closes and reopens at a single momentum, while the order
parameters and diode efficiency $\eta$ evolve smoothly without hysteresis.
This classification is corroborated by the periodic-boundary BdG spectra
in Fig.\,\,\ref{fig:EndMatt}(b) for the three representative points
marked in Fig.\,\,\ref{fig:EndMatt}(a), and by the open-boundary
spectra in Fig.\,\,\ref{fig:EndMatt}(c), which show the emergence
of topological in-gap MZMs exclusively upon entering the tFFLO phase.

\clearpage


\appendix

\setcounter{section}{0}
\setcounter{subsection}{0}
\setcounter{equation}{0}
\setcounter{figure}{0}
\setcounter{table}{0}

\renewcommand{\thesection}{S\arabic{section}}
\renewcommand{\thesubsection}{S\arabic{section}.\arabic{subsection}}
\renewcommand{\theequation}{S\arabic{equation}}
\renewcommand{\thefigure}{S\arabic{figure}}
\renewcommand{\thetable}{S\arabic{table}}

\section*{Supplemental Material for\\
``First-Order Topological FFLO Transition and Superconducting Diode Sign Reversal in Altermagnetic Nanowires''}

\addcontentsline{toc}{section}{Supplemental Material}

In this Supplemental Material, we present details about lattice model
(Sec. S1), mean-field treatment of FFLO pairing and zero-temperature free energy
(Sec. S2), particle--hole symmetry and $\mathbb{Z}_2$ topological invariant
(Sec. S3), phase diagram as a function of chemical potential
(Sec. S4), Ginzburg-Landau analysis (Sec. S5), and temperature-induced crossover
from first- to second-order tFFLO transition (Sec. S6).


\section{Lattice model}

We consider a one-dimensional spin-orbit-coupled nanowire on the surface of a $d$-wave altermagnet.
The normal-state lattice Bloch Hamiltonian is
\begin{equation}
h(k) = t\cos k \,\sigma_0
      + \alpha_y \sin k\,\sigma_y
      + \big(\alpha_z\sin k + J_A \cos k\big)\,\sigma_z,
\end{equation}
with $t$ the hopping amplitude, $\alpha_y$ and $\alpha_z$ the Rashba- and Dresselhaus-like
spin-orbit couplings, and $J_A$ the altermagnetic exchange. Here $\sigma_i$ ($i = x,y,z$) are
Pauli matrices in spin space and $\sigma_0$ is the identity.
We denote the electron spinor as
\begin{equation}
\psi_k =
\begin{pmatrix}
c_{k\uparrow}\\
c_{k\downarrow}
\end{pmatrix},
\end{equation}
so that the normal-state Hamiltonian reads
\begin{equation}
H_0 = \sum_k \psi_k^\dagger \big(h(k) - \mu\big)\psi_k,
\end{equation}
where  $\mu$ is the chemical potential.

To describe a finite-momentum $s$-wave FFLO state, we introduce the Nambu spinor
\begin{equation}
\Phi_k^\dagger =
\big(
\psi_{k+\frac{q}{2}}^\dagger,
\psi_{-k+\frac{q}{2}}^T i\sigma_y
\big),
\end{equation}
and assume a uniform pairing amplitude $\Delta$ at center-of-mass momentum $q$. The
Bogoliubov--de~Gennes (BdG) Hamiltonian can then be written as
\begin{equation}
H = \frac{1}{2} \sum_k \Phi_k^\dagger \mathcal{H}_{\text{k}}(q)\,\Phi_k,
\end{equation}
with
\begin{equation}
\mathcal{H}(k) =
\begin{pmatrix}
h\big(k+\tfrac{q}{2}\big)-\mu & \Delta\,\sigma_0 \\
\Delta\,\sigma_0 & -\sigma_y h^T\big(-k+\tfrac{q}{2}\big)\sigma_y + \mu
\end{pmatrix}.
\label{eq:H_BdG_matrix}
\end{equation}

The real-space form of the pairing term follows from
\begin{align}
&\Delta \sum_k c_{k+\frac{q}{2},\uparrow}^\dagger
                 c_{-k+\frac{q}{2},\downarrow}^\dagger
\nonumber\\
\rightarrow\;
&\Delta\sum_{k}\sum_{n,n'}
c_{n\uparrow}^\dagger c_{n'\downarrow}^\dagger
e^{-i(k+q/2)n}\,e^{-i(-k+q/2)n'}
=\Delta\sum_n e^{-iqn} c_{n\uparrow}^\dagger c_{n\downarrow}^\dagger,
\end{align}
demonstrating the FFLO modulation $e^{-iqn}$ along the wire.
In real space, it is convenient to introduce the Nambu spinor
\begin{equation}
\Psi_n =
\begin{pmatrix}
c_{n\uparrow}\\
c_{n\downarrow}\\
c_{n\downarrow}^\dagger\\
-\,c_{n\uparrow}^\dagger
\end{pmatrix},
\end{equation}
where $\tau_i$ act in Nambu (particle--hole) space. The full tight-binding BdG Hamiltonian reads
\begin{align}
H &= \frac{1}{2}\sum_n \Big[
 \Psi_n^\dagger\big(-\mu \tau_z + \Delta\cos(qn)\tau_x - \Delta\sin(qn)\tau_y\big)\sigma_0 \Psi_n
\nonumber\\
&\quad
+\Psi_n^\dagger
\Big(t\,\tau_z \sigma_0 + J_A\,\tau_0\sigma_z
+ i\,\tau_z(\alpha_y\sigma_y+\alpha_z\sigma_z)\Big)\Psi_{n+1}
+ \text{h.c.}
\Big].
\end{align}
Equivalently, in terms of electron operators,
\begin{align}
H &= \frac{t}{2}\sum_{n,\sigma} c_{n\sigma}^\dagger c_{n+1,\sigma}
+ \frac{\alpha_y}{2}\sum_{n,\sigma} \sigma\, c_{n\sigma}^\dagger c_{n+1,\bar\sigma}
\nonumber\\
&\quad
+ \frac{J_A+i\alpha_z}{2}\sum_{n,\sigma} \sigma\, c_{n\sigma}^\dagger c_{n+1,\sigma}\nonumber\\
&\quad
-\sum_n\Big(\Delta e^{-iqn} c_{n\uparrow}^\dagger c_{n\downarrow}^\dagger + \text{h.c.}\Big)
- \mu\sum_{n,\sigma} c_{n\sigma}^\dagger c_{n\sigma}.
\end{align}
We denote the eigenvalues and eigenvectors of Eq.~\eqref{eq:H_BdG_matrix} as
\begin{equation}
\mathcal{H}(k)\,\varphi_{k,s} = E_{k,s}\,\varphi_{k,s},
\qquad s=1,2,3,4,
\end{equation}
with
\begin{equation}
\varphi_{k,s} =
\big(u_{k\uparrow,s},\,u_{k\downarrow,s},\,v_{k\downarrow,s},\,v_{k\uparrow,s}\big)^T.
\end{equation}

\section{Mean-field treatment of FFLO pairing and zero-temperature free energy}

To generate the finite-momentum $s$-wave pairing self-consistently, we introduce a local
attractive Hubbard interaction
\begin{equation}
H_I = -\frac{U}{N_c}\sum_{k,k',q}
c_{k+q/2,\uparrow}^\dagger c_{-k+q/2,\downarrow}^\dagger
c_{-k'+q/2,\downarrow} c_{k'+q/2,\uparrow},
\end{equation}
where $N_c$ is the number of lattice sites and $U>0$.

We define the $s$-wave FFLO order parameter with finite momentum $q$ as
\begin{equation}
\hat{\Delta}_q = \frac{U}{N_c}\sum_k
c_{-k+q/2,\downarrow} c_{k+q/2,\uparrow},
\end{equation}
so that the interaction term becomes
\begin{equation}
H_I = -\frac{N_c}{U}\sum_q \hat{\Delta}_q^\dagger \hat{\Delta}_q.
\end{equation}
In the mean-field approximation we take
\begin{equation}
\langle \hat{\Delta}_q\rangle = \Delta, \qquad
\langle \hat{\Delta}_q^\dagger\rangle = \Delta^*,
\end{equation}
with $\Delta$ chosen real in what follows. The mean-field Hamiltonian then reads
\begin{align}
\hat{H}_{\text{MF}} - \mu \hat{N}
&= \sum_k \psi_k^\dagger\big(h(k)-\mu\big)\psi_k\nonumber\\
&\quad
- \big(\Delta \sum_k c_{k+q/2,\uparrow}^\dagger c_{-k+q/2,\downarrow}^\dagger
- h.c.\big)
+ \frac{N_c}{U}\Delta^2.
\end{align}
Rewriting in Nambu space and symmetrizing in $k$, one finds
\begin{equation}
\hat{H}_{\text{MF}} - \mu \hat{N}
= \frac{1}{2}\sum_k \Phi_k^\dagger \mathcal{H}(k)\,\Phi_k
- N_c \mu + \frac{N_c}{U}\Delta^2,
\end{equation}
where $\mathcal{H}_{\text{BdG}}(k)$ is the $4\times4$ matrix given in
Eq.~\eqref{eq:H_BdG_matrix}. This establishes the connection between the microscopic interaction
and the phenomenological BdG description.

Diagonalizing the BdG Hamiltonian, we introduce quasiparticle operators
\begin{align}
\gamma_{k,s}^\dagger
= &\;
u_{k\uparrow,s} c_{k+q/2,\uparrow}^\dagger
+ u_{k\downarrow,s} c_{k+q/2,\downarrow}^\dagger
\nonumber\\
&\quad- v_{k\downarrow,s} c_{-k+q/2,\downarrow}
+ v_{k\uparrow,s} c_{-k+q/2,\uparrow},
\end{align}
such that
\begin{equation}
H = \frac{1}{2}\sum_{k,s} E_{k,s} \gamma_{k,s}^\dagger \gamma_{k,s}
+ \frac{N_c}{U}\Delta^2 - N_c \mu.
\end{equation}
At zero temperature, the free-energy density is
\begin{align}
F(\Delta,q)
&= \frac{1}{N_c}\langle \hat{H}_{\text{MF}} - \mu \hat{N}\rangle
\nonumber\\
&= \frac{1}{2N_c}\sum_{k,s} E_{k,s} \Theta(-E_{k,s})
+ \frac{\Delta^2}{U} - \mu,
\end{align}
which in the continuum limit becomes
\begin{equation}
F(\Delta,q)
= \frac{1}{2}\int_{-\pi}^{\pi} \frac{dk}{2\pi}
\sum_{s=1}^{4} E_{k,s}\,\Theta(-E_{k,s})
+ \frac{\Delta^2}{U} - \mu.
\end{equation}
The self-consistent order parameter $\Delta$ for a given $q$ is obtained by imposing
\begin{equation}
\frac{\partial F(\Delta,q)}{\partial \Delta} = 0,
\end{equation}
which leads to
\begin{equation}
\frac{2\Delta}{U} =
-\frac{1}{2}\int_{-\pi}^{\pi} \frac{dk}{2\pi}
\sum_{s=1}^{4}
\frac{\partial E_{k,s}}{\partial \Delta}\,\Theta(-E_{k,s}).
\end{equation}
In practice we solve this equation numerically for each $q$ to obtain the optimal
$(\Delta,q)$ profile and identify first-order transitions between distinct FFLO
phases as discussed in the main text.

\section{Particle--hole symmetry and $\mathbb{Z}_2$ topological invariant}

By construction, the BdG Hamiltonian belongs to symmetry class D and obeys the antiunitary
particle--hole (PH) symmetry
\begin{equation}
\mathcal{P}\,\mathcal{H}^{*}(k)\,\mathcal{P}^{-1} = -\mathcal{H}(-k),\qquad
\mathcal{P}=\tau_x,
\end{equation}
where $\tau_x$ acts in Nambu space. This antiunitary symmetry guarantees that the spectrum
comes in pairs
$
E_s(k),\ -E_{s^\prime}(-k).
$
In 1D, symmetry class D is characterized by a $\mathbb{Z}_2$ topological
invariant, which we compute using the Pfaffian formalism.

To bring the Hamiltonian into a manifestly antisymmetric form at the particle--hole symmetric
points $k=0,\pi$, we perform the unitary transformation
\begin{equation}
U = \frac{1}{\sqrt{2}}(1 + i\tau_x),\qquad
U^\dagger = \frac{1}{\sqrt{2}}(1 - i\tau_x),
\end{equation}
and define
\begin{equation}
H'(k) = U^\dagger \mathcal{H}(k) U.
\end{equation}
Using $\tau_x \mathcal{H}^{T}(k) \tau_x = -\mathcal{H}(-k)$, one finds
\begin{equation}
\big(U^\dagger \mathcal{H}(k)\, U\big)^{T}
= -U^\dagger \mathcal{H}(-k)\, U,
\end{equation}
so that $H'(k)$ is antisymmetric at the time-reversal-invariant momenta (TRIM) $k=0,\pi$.

Evaluating $H'(k)$ at $k=0$ and $k=\pi$, and using the compact expressions derived in the
main text, the Pfaffians at the particle--hole symmetric momenta are
\begin{widetext}
\begin{align}
\operatorname{Pf}\big(H'(0)\big)
&=
\big(\mu - t\cos\tfrac{q}{2}\big)^2
- \big(J_A\cos\tfrac{q}{2} + \alpha_z\sin\tfrac{q}{2}\big)^2
+ \Delta^2 - \alpha_y^2 \sin^2\!\Big(\frac{q}{2}\Big),
\label{eq:Pf0}
\\[6pt]
\operatorname{Pf}\big(H'(\pi)\big)
&=
\big(\mu + t\cos\tfrac{q}{2}\big)^2
- \big(J_A\cos\tfrac{q}{2} + \alpha_z\sin\tfrac{q}{2}\big)^2
+ \Delta^2 - \alpha_y^2 \sin^2\!\Big(\frac{q}{2}\Big).
\label{eq:Pfpi}
\end{align}
\end{widetext}

The $\mathbb{Z}_2$ topological index is then given by
\begin{equation}
\nu = \operatorname{sgn}\big[\operatorname{Pf}\big(H'(0)\big)\big]\,
\operatorname{sgn}\big[\operatorname{Pf}\big(H'(\pi)\big)\big].
\end{equation}
A topological phase with Majorana end modes corresponds to $\nu=-1$, which requires that the
Pfaffians at $k=0$ and $k=\pi$ have opposite signs:
\begin{equation}
\operatorname{Pf}\big(H'(0)\big)\,
\operatorname{Pf}\big(H'(\pi)\big) < 0.
\end{equation}

From Eqs.~\eqref{eq:Pf0} and \eqref{eq:Pfpi}, the condition for the existence of a topological
phase can be cast in the form
\begin{equation}
\Big(J_A\cos\!\frac{q}{2} + \alpha_z\sin\!\frac{q}{2}\Big)^2
+ \alpha_y^2\sin^2\!\Big(\frac{q}{2}\Big) > \Delta^2,
\end{equation}
and the corresponding chemical potential window
\begin{widetext}
\begin{equation}
t\cos\!\Big(\frac{q}{2}\Big)
-\sqrt{\Big(J_A\cos\!\frac{q}{2} + \alpha_z\sin\!\frac{q}{2}\Big)^2
+ \alpha_y^2\sin^2\!\Big(\frac{q}{2}\Big) - \Delta^2}
< \mu
<
t\cos\!\Big(\frac{q}{2}\Big)
+\sqrt{\Big(J_A\cos\!\frac{q}{2} + \alpha_z\sin\!\frac{q}{2}\Big)^2
+ \alpha_y^2\sin^2\!\Big(\frac{q}{2}\Big) - \Delta^2}.
\end{equation}
\end{widetext}
Within this interval of $\mu$, the system is in the topological
FFLO phase discussed in the main text, featuring Majorana bound states.

\begin{figure}[t]
    \centering
    \includegraphics[width=0.8\linewidth]{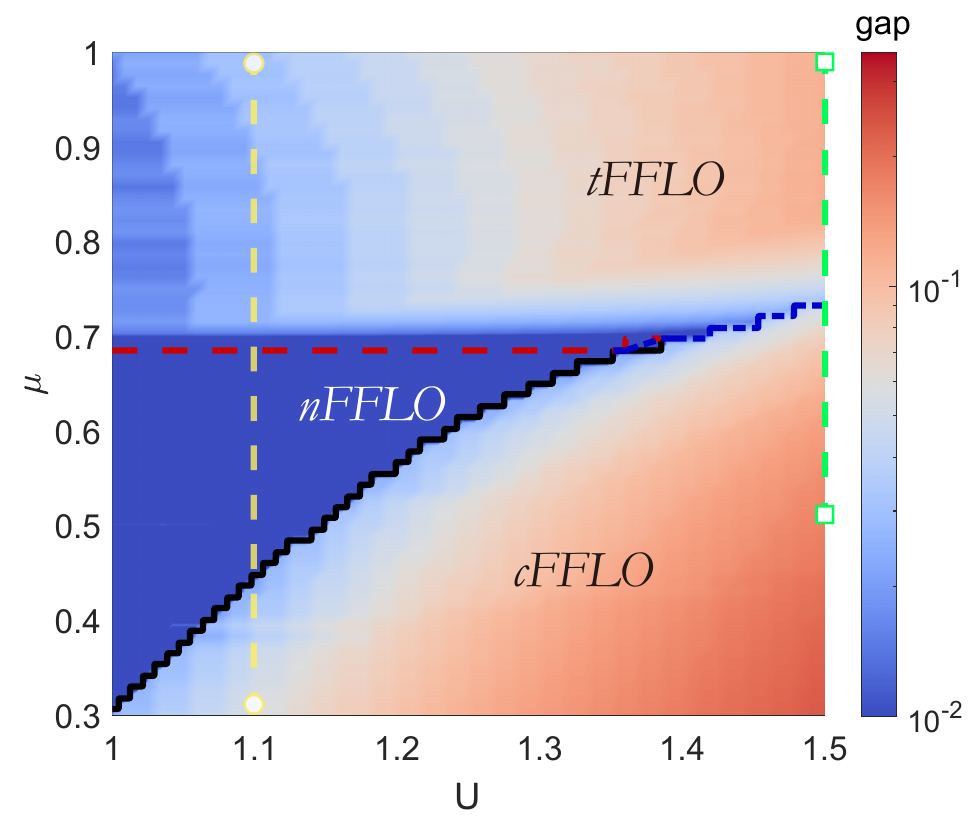}
    \caption{Zero-temperature phase diagram in the $(U,\mu)$ plane at fixed altermagnetic exchange $J_A=0.4$.
        The color scale represents the quasiparticle excitation gap on a logarithmic scale, and the three FFLO phases---conventional FFLO (cFFLO), nodal FFLO (nFFLO), and topological FFLO (tFFLO)---are labeled in their respective regions.
        The thick black and blue curves denote first-order phase boundaries, where the optimal pairing parameters $(\Delta_{\text{opt}},q_{\text{opt}})$ exhibit discontinuous jumps.
        The red dashed curve marks a continuous (second-order) transition where the gap closes smoothly at isolated nodal points.
        Vertical yellow and green dashed lines indicate representative cuts at fixed interaction strengths $U=1.1$ and $U=1.5$, respectively; the corresponding evolution of the order parameters, energy spectra, and $\mathbb{Z}_2$ topological index along these cuts is shown in Fig.~\ref{fig:S1}.
    }
    \label{fig:S2}
\end{figure}
\section{Phase diagram as a function of chemical potential}

\begin{figure}[t]
    \centering
    \includegraphics[width=0.8\linewidth]{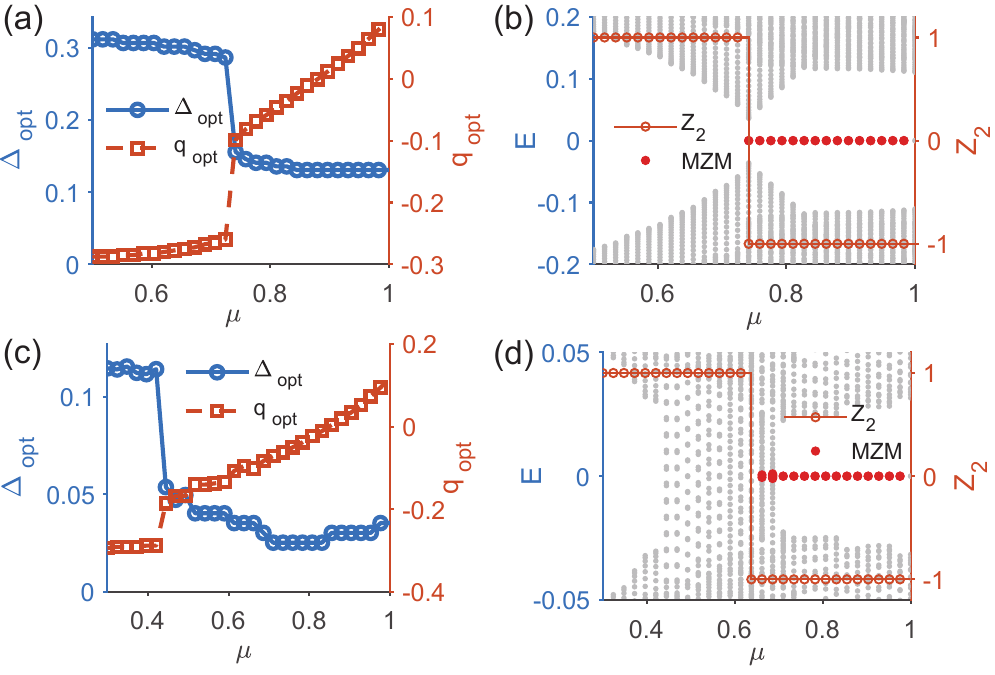}
    \caption{ Phase evolution of the one-dimensional FFLO superconductor as a function of chemical potential $\mu$ for two interaction strengths. 
        (a,b) For $U=1.5$, the system exhibits a direct transition from a conventional FFLO (cFFLO) phase to a topological FFLO (tFFLO) phase. 
        Panel (a) shows the optimal pairing amplitude $\Delta_{\text{opt}}$ (left axis) and center-of-mass momentum $q_{\text{opt}}$ (right axis) versus $\mu$. 
        Panel (b) displays the open-boundary energy spectrum $E$ (left axis, gray dots) and the topological invariant $\mathbb{Z}_{2}$ (right axis, orange line) as functions of $\mu$; zero-energy Majorana bound states (MZMs) in the topological regime ($\mathbb{Z}_{2}=-1$) are highlighted by red symbols. 
        (c,d) For $U=1.1$, the system undergoes a sequence of cFFLO–nFFLO–tFFLO transitions upon increasing $\mu$. 
        Panel (c) shows the corresponding evolution of $\Delta_{opt}$ and $q_{opt}$, while panel (d) presents the open-boundary spectrum and $\mathbb{Z}_{2}$, again marking the emergence of zero-energy MZMs in the tFFLO phase.}
    \label{fig:S1}
\end{figure}

In the main text, we focused on the phase structure of the FFLO state as a function of the
altermagnetic exchange $J_A$ at fixed chemical potential $\mu$. Here we demonstrate that
varying $\mu$ plays a qualitatively similar role and can also drive a sequence of
conventional, nodal, and topological FFLO phases.

Figure~\ref{fig:S2} displays the zero-temperature phase diagram in the $(U,\mu)$ plane for a
fixed $J_A=0.4$. The color scale indicates the quasiparticle excitation gap (on a logarithmic
scale), and three distinct FFLO phases are identified: a conventional gapped FFLO phase
(cFFLO), an intermediate nodal FFLO phase (nFFLO) with bulk gapless excitations, and a
topological FFLO phase (tFFLO) hosting Majorana zero modes. The thick black and blue curves
separate the cFFLO and nFFLO regions and the nFFLO and tFFLO regions via first-order
transitions, signaled by discontinuous jumps of the optimal pairing amplitude
$\Delta_{\text{opt}}$ and center-of-mass momentum $q_{\text{opt}}$. In contrast, the red dashed line indicates a continuous (second-order) phase boundary: the Bogoliubov quasiparticle gap opens continuously as the system evolves from the nFFLO to the tFFLO phase, while the order parameters change smoothly without any discontinuity.

To elucidate the structure of these transitions, Fig.~\ref{fig:S1} shows one-dimensional cuts
through the phase diagram of Fig.~\ref{fig:S2} at fixed $U=1.1$ (yellow dashed line) and
$U=1.5$ (green dashed line). For $U=1.1$, increasing $\mu$ drives a sequence of
cFFLO--nFFLO--tFFLO phases, with corresponding changes in $(\Delta_{\text{opt}},q_{\text{opt}})$,
the open-boundary spectrum, and the $\mathbb{Z}_2$ topological index. For $U=1.5$, the nodal
region is suppressed and the system undergoes a direct first-order transition from cFFLO to
tFFLO upon tuning $\mu$. These results demonstrate that, in close analogy to the $J_A$-driven
phase diagrams discussed in the main text, the chemical potential can also be used as an
effective tuning knob to realize and control first-order and second-order transitions between
cFFLO, nFFLO, and tFFLO phases.

\section{Ginzburg-Landau analysis}

In the main text we qualitatively used a Ginzburg-Landau (GL) picture to
illustrate the first-order tFFLO–cFFLO transition and the associated
superconducting diode effect. Here we collect the technical details that
are \emph{not} contained there, focusing on: (i) explicit analytic
expressions for the competing GL branches, (ii) the conditions for a
first-order versus continuous transition, and (iii) how the double-valley
structure controls the diode response.

We start from a phenomenological GL free energy for the FFLO order
parameter,
\begin{equation}
    \Omega(q,\Delta)
    = \alpha(q)\,\Delta^{2}
      + \beta(q)\,\Delta^{4}
      + \gamma(q)\,\Delta^{6},
    \label{eq:GL-F-supp}
\end{equation}
with $\gamma(q)>0$ ensuring stability at large $|\Delta|$. A first-order
transition requires $\beta(q)<0$ in a window of $q$, which generates a
barrier between two superconducting minima.

Imposing the stationarity condition $\partial \Omega/\partial\Delta=0$ gives
\begin{equation}
    2\alpha(q)\Delta
    + 4\beta(q)\Delta^{3}
    + 6\gamma(q)\Delta^{5} = 0,
\end{equation}
so that in addition to the normal state $\Delta=0$ there are two nontrivial
branches
\begin{equation}
    \Delta_{\pm}^{2}(q)
    = \frac{-\beta(q) \pm \sqrt{\beta(q)^{2} - 3\alpha(q)\gamma(q)}}{3\gamma(q)}.
    \label{eq:Delta-pm-supp}
\end{equation}
These correspond to two SC valleys in the $(\Delta,q)$ free-energy
landscape. Substituting \eqref{eq:Delta-pm-supp} back into
\eqref{eq:GL-F-supp} yields two effective free-energy branches
\begin{equation}
    \Omega_{\pm}(q) \equiv \Omega\bigl(q,\Delta_{\pm}(q)\bigr),
\end{equation}
which can be written in closed form as
\begin{equation}
    \Omega_{\pm}(q)
    = \pm\,\frac{
       \bigl[\sqrt{\beta^{2}-3\alpha\gamma} \mp \beta\bigr]\,
       \Bigl(6\alpha\gamma
             \pm \beta\bigl[\sqrt{\beta^{2}-3\alpha\gamma}
                               \mp \beta\bigr]\Bigr)}
       {27\gamma^{2}}.
    \label{eq:F-pm-exact-supp}
\end{equation}
In the main text, we used only their qualitative structure; here we provide
the full expressions for completeness.

In the parameter regimes relevant to our microscopic model, $\Omega_{\pm}(q)$
are accurately approximated by simple quadratics,
\begin{equation}
    \Omega_{\pm}(q) \approx f_{\pm0} + f_{\pm1}\,q + f_{\pm2}\,q^{2},
    \label{eq:Fpm-quad-supp}
\end{equation}
with coefficients $f_{\pm i}$ obtained by fitting to the numerical
free-energy data. This quadratic reduction is what underlies the nearly
parabolic valleys shown in Figs. 3(c,d) in the main text.

Equation~\eqref{eq:Delta-pm-supp} shows that nonzero solutions exist
whenever
\begin{equation}
    \beta(q)^{2} > 3\alpha(q)\gamma(q).
\end{equation}
The line $\beta(q)^{2} = 3\alpha(q)\gamma(q)$ corresponds to the point
where the two nontrivial branches $\Delta_{+}$ and $\Delta_{-}$ merge and
disappear. The nature of the SC transition at given $q$ is controlled by
the sign of $\beta(q)$:

\begin{itemize}
    \item For $\beta(q)>0$, the transition from $\Delta=0$ to a single
          nonzero minimum is continuous (second order) as $\alpha(q)$
          changes sign.
    \item For $\beta(q)<0$, two finite-$\Delta$ minima coexist with the
          normal state in a range of $\alpha(q)$, and the system can jump
          between them at fixed $q$ as a control parameter is tuned, giving
          a first-order SC transition.
\end{itemize}

In our microscopic model, the tFFLO-cFFLO transition line lies in the
regime where $\beta(q)<0$ near the relevant $q$ values, so that the
competition between the two SC valleys is intrinsically first order.

\begin{figure}[t]
    \centering
    \includegraphics[width=0.5\columnwidth]{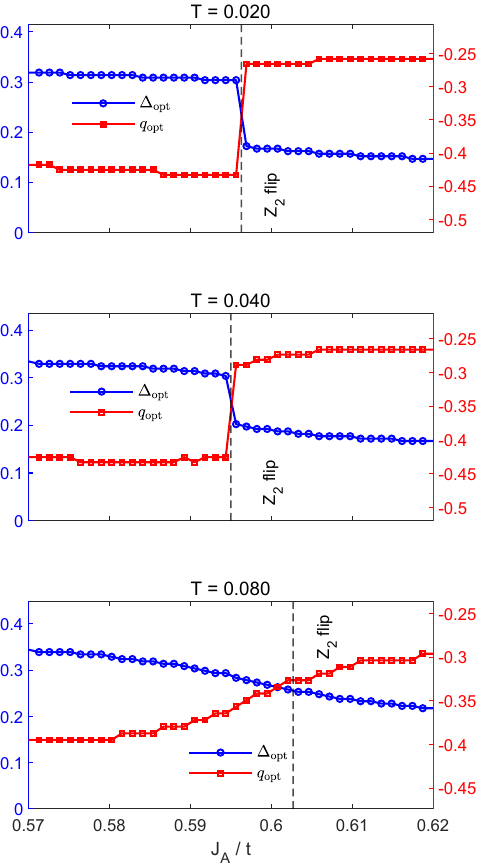}
    \caption{Temperature-induced change of the cFFLO–tFFLO transition order as a function of altermagnetic exchange $J_A/t$.
    For each temperature $T/t = 0.020,\,0.040,\,0.080$ (top to bottom), we plot the optimal pairing amplitude $\Delta_{\mathrm{opt}}$ (left axis, blue) and center-of-mass momentum $q_{\mathrm{opt}}$ (right axis, red) obtained by minimizing the free energy $\Omega(\Delta,q)$.
    The vertical dashed line in each panel marks the phase boundary determined by the sign change of the $\mathbb{Z}_2$ invariant evaluated at $(\Delta_{\mathrm{opt}},q_{\mathrm{opt}})$.
    At low temperatures $T/t=0.020$ and $0.040$, both $\Delta_{\mathrm{opt}}$ and $q_{\mathrm{opt}}$ exhibit clear discontinuous jumps at the transition, characteristic of a first-order cFFLO–tFFLO transition.
    At higher temperature $T/t=0.080$, $\Delta_{\mathrm{opt}}$ and $q_{\mathrm{opt}}$ instead evolve smoothly through the $\mathbb{Z}_2$ sign flip, indicating that the transition has become continuous in the order parameters.
    This behavior reflects the thermal softening of the double-valley free-energy structure that drives the first-order switch at low $T$.}
    \label{fig:T_first_second}
\end{figure}

The key feature that is not fully elaborated in the main text is how the
\emph{double-valley} structure in $\Omega_{\pm}(q)$ feeds into the
superconducting diode effect. Along each branch, the supercurrent follows
from the current–phase relation
\begin{equation}
    J_{\pm}(q) = 2\frac{\partial \Omega_{\pm}(q)}{\partial q}
               \approx 2f_{\pm1} + 4 f_{\pm2}\,q,
    \label{eq:Jpm-supp}
\end{equation}
and thus varies approximately linearly with $q$ within that valley.

We denote by $q_c$ the momentum at which the global minimum switches from
the $\Omega_{+}$ valley to the $\Omega_{-}$ valley, and by $q_{\pm}$ the endpoint
momenta at which the corresponding SC solution vanishes,
$\Omega_{\pm}(q_{\pm})=0$. The critical currents in each branch are then
obtained by evaluating $J_{\pm}(q)$ at the relevant endpoints, and the
branch-resolved diode efficiencies can be written in a compact form as
\begin{equation}
    \eta_{\pm}
    = \pm\,\frac{f_{\pm1}/f_{\pm2} + (q_{c}+q_{\pm})}{q_{c} - q_{\pm}},
    \label{eq:eta-pm-supp}
\end{equation}
as quoted in the main text. The important point here is that the sign and
magnitude of $\eta_{\pm}$ are controlled by (i) the slope and curvature
$f_{\pm1},f_{\pm2}$ of each valley, and (ii) the relative position of the
switching point $q_c$ and the extinction momenta $q_{\pm}$.

Near the first-order tFFLO–cFFLO boundary, the free energy
$\Omega(q,\Delta)$ exhibits two nearly degenerate minima at distinct
finite momenta, leading to a ``double-valley'' structure in the reduced
$\Omega_{\pm}(q)$. Small changes in the external control parameter (e.g.~$J_A$)
can move the system across the degeneracy condition
$\Omega_{+}^{\mathrm{ex}}=\Omega_{-}^{\mathrm{ex}}$, causing the ground state to jump
between the two valleys and thereby changing the sign of the dominant
critical current. This mechanism, rooted purely in the geometry of the two
GL branches, explains why the superconducting diode efficiency $\eta$ is so
sensitive near the first-order line and why it exhibits an abrupt sign
reversal across the transition.

These results show that the numerically observed abrupt changes in the
pairing strength and diode efficiency are natural consequences of a
first-order transition between two competing FFLO valleys in the GL free
energy.

\section{Temperature-induced crossover from first- to second-order tFFLO transition}

So far we have focused on the zero-temperature phase diagram, where the transition between the conventional FFLO (cFFLO) and topological FFLO (tFFLO) states is sharply first order. We now show that finite temperature can qualitatively modify the nature of this transition, driving a crossover from first-order to continuous (second-order-like) behavior.

To this end, we fix the microscopic parameters in the FFLO regime and scan the altermagnetic exchange $J_A/t \in [0.57,0.62]$ at several temperatures $T/t$, extracting at each point the optimal pairing amplitude $\Delta_{\mathrm{opt}}$ and center-of-mass momentum $q_{\mathrm{opt}}$ by minimizing the free energy $\Omega(\Delta,q)$. In parallel, we compute the $\mathbb{Z}_2$ topological invariant evaluated at the same $(\Delta_{\mathrm{opt}},q_{\mathrm{opt}})$, and use the sign change of the invariant (from $+1$ to $-1$) to locate the cFFLO–tFFLO phase boundary.

Figure~\ref{fig:T_first_second} summarizes the results for three representative temperatures $T/t = 0.020,\,0.040,\,0.080$. At the lowest temperatures [top and middle panels of Fig.~\ref{fig:T_first_second}], the transition between cFFLO and tFFLO is strongly first order: as $J_A/t$ is tuned through the critical value marked by the dashed line (determined by the $\mathbb{Z}_2$ sign flip), both $\Delta_{\mathrm{opt}}$ and $q_{\mathrm{opt}}$ exhibit clear discontinuous jumps. The simultaneous nonanalytic change in both the pairing strength and center-of-mass momentum is the hallmark of a first-order transition between two competing FFLO configurations.

By contrast, at higher temperature $T/t=0.080$ [bottom panel of Fig.~\ref{fig:T_first_second}], the same cFFLO–tFFLO boundary—again identified unambiguously by the $\mathbb{Z}_2$ sign change—no longer coincides with discontinuities in $\Delta_{\mathrm{opt}}$ or $q_{\mathrm{opt}}$. Instead, both quantities evolve smoothly across the transition point, indicating that thermal fluctuations have softened the underlying free-energy landscape: the double-valley structure that drives the first-order switch at low $T$ gradually merges into a single effective minimum at higher $T$, so that the cFFLO–tFFLO transition becomes continuous in the order parameters even though it still separates distinct topological phases.

Taken together, these results demonstrate a temperature-driven crossover from a sharply first-order topological transition at low $T$ to a second-order-like, continuous transition at elevated $T$. The phase boundary itself remains topologically protected—marked by the $\mathbb{Z}_2$ sign change—but the thermodynamic signatures in $\Delta_{\mathrm{opt}}$ and $q_{\mathrm{opt}}$ evolve from discontinuous to smooth as thermal fluctuations wash out the free-energy barrier between the two FFLO valleys.

\end{document}